\documentclass[12pt,preprint]{aastex}

\begin{document}

\title{Radio Properties of Low Redshift Broad Line Active Galactic Nuclei Including Extended Radio Sources}

\author{Stephen E. Rafter\altaffilmark{1}, D. Michael Crenshaw, Paul J. Wiita\altaffilmark{2}}
\affil{Department of Physics \& Astronomy, Georgia State University, Atlanta, GA 30303}
\altaffiltext{1}{Physics Department, the Technion, Haifa 32000, Israel; e-mail: rafter@physics.technion.ac.il}
\altaffiltext{2}{Department of  Physics, The College of New Jersey, Ewing, NJ 08628}
\begin{abstract}
\indent

We present a study of the extended radio emission in a sample of 8434 low
redshift (z $<$ 0.35) broad line active galactic nuclei (AGN) from the Sloan
Digital Sky Survey (SDSS). To calculate the jet and lobe contributions to the
total radio luminosity, we have taken the 846 radio core sources detected in our
previous study of this sample and performed a systematic search in the Faint
Images of the Radio Sky at Twenty-centimeters (FIRST) database for extended
radio emission that is likely associated with the optical counterparts. We found
51 out of 846 radio core sources have extended emission ($>$ 4$''$ from the
optical AGN) that is positively associated with the AGN, and we have
identified an additional 12 AGN with extended radio emission but no detectable
radio core emission. Among these 63 AGN, we found 6 giant radio galaxies (GRGs), with
projected emission exceeding 750 kpc in length, and several other AGN with
unusual radio morphologies also seen in higher redshift surveys.  The optical
spectra of many of the extended sources are similar to that of typical broad
line radio galaxy spectra, having broad H$\alpha$ emission lines with boxy
profiles and large M$_{\rm BH}$.  With extended emission taken into account, we
find strong evidence for a bimodal distribution in the
radio-loudness parameter $\mathcal{R}$ ($\equiv \nu_{\rm radio}$L$_{\rm
radio}/\nu_{\rm opt}$L$_{\rm opt}$), where the lower radio luminosity
core-only sources appear as a population separate from the extended sources, with
a dividing line at log($\mathcal{R}$) $\approx 1.75$. This dividing line ensures
that these are indeed the most radio-loud AGN, which may have different or
extreme physical conditions in their central engines when compared to the more
numerous radio quiet AGN.

\end{abstract}


~~~~~.

\section{Introduction}
\indent

It is not uncommon to find a particularly radio luminous active galactic nucleus
(AGN) classified as a broad line radio galaxy (BLRG), a quasar, \it and \rm  a
\citet{1974MNRAS.167P..31F} class II object (an FR II's radio emission is
lobe-dominated and edge-brightened, whereas an FR I is jet-dominated and
edge-darkened).  Generally, an AGN's classification can depend on many factors
such as when and in what part of the spectrum it was first discovered, which
particular study it is being used in, and the source of the data.  Moving beyond
an often blurred and overlapping system of identification into one based on more
quantitative parameters could allow a more continuous classification scheme that
is easier to apply to the large samples that continue to become available with
large area surveys, e.g., in the radio (VLA's FIRST survey), infrared (2MASS),
optical (SDSS), and X-ray (ROSAT) bands. The AGN in these large samples can now
be classified based on measured quantities in a statistical fashion that is
inherently more continuous than the discrete nomenclature generally used
\citep[e.g.,][]{2006MNRAS.372..961K}.

All AGN are believed to be powered by the accretion of matter onto a
supermassive black hole (SMBH), and show strong emission and variability in all
wavebands from the radio to X-ray regimes.  Although not a fundamental physical
parameter, the inclination of the BH/accretion disk system to our line of sight
is an observational parameter that, according to accepted unification models, is
responsible for the presence or absence of permitted broad lines (BL) (in type 1
and type 2 AGN, respectively) in optical spectra due to toroidal obscuration by
gas and dust when viewed at large inclination angles \citep{1993ARA&A..31..473A,
1995PASP..107..803U}.  Typically we assume that BL AGN have lower inclinations
to our line of sight and we are looking down onto the BL region (BLR) clouds
that lie just outside the immediate vicinity of the BH and accretion disk
system.

An important fundamental parameter is black hole mass (M$_{\rm BH}$), which has
been determined directly through various methods including H$_{\rm 2}$O-maser
observations and reverberation mapping (RM) \citep[e.g.,][]{1995PNAS...9211427M,
Peterson2004}.  RM becomes a powerful tool when applied to large spectroscopic
samples in that the scaling relations derived from RM analysis allow single
epoch (single spectra) M$_{\rm BH}$ determinations for BL AGN
\citep[e.g.,][]{Kaspi2000, Kaspi2005, 2009ApJ...697..160B}.  Another important
parameter, the Eddington ratio, is the ratio of the bolometric luminosity
(L$_{\rm bol}$) to the Eddington luminosity (L$_{\rm Edd}$ $\propto$ M$_{\rm
BH}$).  Determination of the true L$_{\rm bol}$ gives a measure of the accretion
rate and requires a full spectral energy distribution (SED) that spans from
radio to X-ray emission and beyond for most AGN.  AGN that have an observed SED
that spans the entire spectrum provide normalization relations so that one can
use a single continuum measurement in the optical or X-ray bands to stand in as
a reasonable proxy for L$_{\rm bol}$ \citep[e.g.,][]{1994ApJS...95....1E,
2002ApJ...565L..75E, 2004MNRAS.352.1390M}.

The degree of radio-loudness is another means by which to classify AGN, and is
based on the amount of radio emission in the form of core emission, jets and/or
lobes that can be positively associated with the central engine and accretion
phenomenon.  There are two main characterizations of the radio-loudness of AGN.
The first is to set a dividing line between radio-loud (RL) and radio-quiet (RQ)
based on the $\mathcal{R}$ parameter defined as the ratio of the monochromatic 5
GHz radio luminosity to the 4400 \AA ~optical luminosity ($\nu_{\rm 5
GHz}$L$_{\rm 5 GHz}$/$\nu_{\rm 4400}$L$_{\rm 4400}$).  By convention, RL AGN
have $\mathcal{R} >$ 10 and RQ AGN have $\mathcal{R} <$ 10
\citep{Kellermann1989}.  The second way to characterize the degree of
radio-loudness is by using the radio luminosity alone.
\citet{1974MNRAS.167P..31F} originally found a transition from the FR I type
radio morphology to the FR II type corresponding to a luminosity of 10$^{24.5}$
Watts Hz$^{-1}$ at 1.4 GHz \citep{Kawakatu2009}.  While this luminosity is not a
RL/RQ dividing line, the distribution in the radio luminosity plane shows that
most FR Is have luminosities below this dividing line and FR IIs have
luminosities above.  However it is well established that many FR Is are RL when
following the classic $\mathcal{R}$ convention.  Therefore a lower luminosity
dividing line has occasionally been used as an alternate way to classify AGN as
either RL or RQ; e.g., \citet{Best2005} specify 10$^{23}$ Watts Hz$^{-1}$ to be
this division for FIRST data at 1.4 GHz.

A quasar radio dichotomy has been postulated because only 5\% -- 10\% of all AGN
are RL according to the $\mathcal{R} >$ 10 criterion \citep{Kellermann1989,
1995PASP..107..803U, Ivezic2002, 2007ApJ...654...99W}.  This has led to claims
that there is a bimodal distribution in the $\mathcal{R}$ parameter for high
optical luminosity, high redshift sources \citep[ and references
therein]{2003astro.ph.12417L,Ivezic2004}, where usually only the core radio
emission is taken into account.  While the RL AGN are usually thought to be
powered by the same phenomenon of matter accreting onto a SMBH, it has been
suggested that they may have a different accretion mode, e.g., advection
dominated accretion flow \citep{2008NewAR..51..733N} versus a standard thin disk, or that their BH's are more massive or spinning faster, or some combination of both \citep[e.g.,][ and references therein]{Sikora2007}.  Other models propose that powerful jets tap
the spin energy of the BH \citep[e.g.,][]{1977MNRAS.179..433B} so the accretion
rate is nearly irrelevant.  Either case suggests that $\mathcal{R}$, although
not a fundamental quantity, may be linked to one.  Very often the most extreme
RL AGN are FR II types that have giant radio lobes that grow and extend from the
host galaxy out to Mpc scales while being fed by highly collimated jets.
Statistically, these AGN are associated most often with giant elliptical
galaxies that tend to have optical spectra with very broad Balmer line
(H$\alpha$, H$\beta$) profiles with a large full widths at half maximum (FWHM),
typically $>$ 8000 km s$^{-1}$ \citep{Osterbrock2006agnbook}.  In studies of
high redshift, high luminosity AGN, it has generally been thought that most RL
AGN have M$_{\rm BH} > 10^{8}$ M$_{\odot}$ \citep[e.g.,][]{Laor2000,
McLure2004b}.  This clearly manifests itself for most FR IIs when determining
M$_{\rm BH}$ from single epoch measurements, since generically, M$_{\rm BH}
\propto$ FWHM$_{\rm H\alpha}^{2}$L$^{0.5}$ \citep[e.g.,][]{2009ApJ...697..160B}.

It has been shown in studies by \citet{2002ApJ...564..120H} and
\citet{Sikora2007} that there is a strong correlation between radio-loudness and
Eddington ratio, where AGN with very low accretion rates (corresponding to
$\sim$ 10$^{-5}$ L$_{\rm bol}/$L$_{\rm Edd}$) are almost exclusively all RL
based on the $\mathcal{R}$ parameter, and a clear trend can be seen of
decreasing radio-loudness with increasing L$_{\rm bol}/$L$_{\rm Edd}$.  Further,
\citet{Sikora2007} find two separate populations of AGN in the $\mathcal{R}$ vs
L$_{\rm bol}/$L$_{\rm Edd}$ plane, where the upper population consists of FR Is,
BLRGs and RL quasars hosted by giant elliptical galaxies, and the lower
population are mostly Seyfert and Low Ionization Nuclear Emission-line Region
(LINER) types hosted by spiral galaxies.  While these studies do show a
dependence of radio-loudness on accretion rate, they do not exclude the
possibility that there may be other factors that contribute to the generation of
strong radio emission, such as accretion modes which are directly related to the
amount of matter in the accretion disk, or the spin of the SMBH.

In \citet{Rafter2009} (hereafter Paper I), we investigated these issues with the
low-redshift sample of broad line AGN from Greene \& Ho (2007), which was
not selected on the basis of any radio property. We found no clear gap between
RL and RQ AGN, and provided evidence for a significant
radio-intermediate population in the local Universe. Using the above
definition, we found that 4.7\% of the AGN in a flux-limited subsample were
radio loud ($\mathcal{R} >$ 10). We also found evidence that the radio-loud
fraction (RLF) decreases with Eddington ratio, in agreement with the above
findings. Finally, we found a significant number of RL AGN with M$_{\rm BH} <
10^{8}$ M$_{\odot}$, which indicates that RL AGN are not a product of only the
most massive black holes in the Universe.

In this paper, we reexamine our sample to study the extended radio emission ($>$
4$''$ from the optical AGN). We investigate the FR I/FR II luminosity
break and its relation to the claimed bimodal distribution in radio loudness.
We also identify a number of unusual radio morphologies for future detailed
study. Finally, we compare these new results to those in Paper I, where only the
core emission was taken into account when calculating L$_{\rm radio}$.

\section{Data Sample and Analysis}
\indent

As discussed in detail in Paper I, we have taken the BL AGN sample from
\citet{Greene2007}, who calculated M$_{\rm BH}$ and L$_{\rm bol}$/L$_{\rm Edd}$
from the full width at half maximum of the broad H$\alpha$ line (FWHM$_{\rm
H\alpha}$) and the luminosity of H$\alpha$ (L$_{\rm H\alpha}$) for 8434 BL AGN
from the Sloan Digital Sky Survey (SDSS) Data Release 4 (DR4).  We performed a
follow up search for these objects using the 2008 April version of the Very
Large Array's (VLA) 1.4 GHz Faint Images of the Radio Sky at Twenty-centimeters
(FIRST) survey database \citep{Becker1995}.  The FIRST survey operates at a
frequency of 1.4 GHz, has an angular resolution of $\sim$ 5$''$ and a limiting
magnitude of 1 mJy.  \citet{Ivezic2002} find that 90\% of all SDSS/FIRST matched
AGN only show radio emission at, or close to, the optical source using a
$1\farcs5$ search radius.  Many of these `core' sources only appear to be
compact or unresolved, meaning that detailed structure on scales $<$ 5$''$ will
not be resolved due to the modest angular resolution of the FIRST survey.  Such
unresolved cores could be arising from orientation effects, where the radio jet
is closely aligned with our line of sight, or they could be fairly young radio
sources whose emission has not had time to expand out to a significant distance
from the host galaxy.  Therefore AGN with unresolved features, whether FR I or
FR II type, will appear as core-only sources in this sample.  

This paper follows our earlier paper in which we used a 4$''$ search radius to
identify radio sources associated with the optical AGN in this sample and where
we showed this leads to very few false radio detections with an optical
counterpart.    We find that of the 8434 objects, 846 have core radio sources
inside this radius (we note that this number is updated from Paper I, where we
found 832 objects using the 2003 April 11 version of the FIRST catalogue).  In
the study of Paper I, only the core radio emission was taken into account in
order to compare that work with other studies (mentioned above) at higher
redshift.  In this work, we have first taken these AGN with radio core emission
and performed a search around a much wider, 60$''$ radius, to identify any
extended emission (at positions $>$ 4$''$ from the optical AGN) that may be
associated with them.  The 60$''$ search radius was chosen due to the fact that
the largest known FR IIs are on the scale of a Mpc \citep[ and references
therein]{2009ApJ...695..156S}, and at the sample redshift limit of $z = 0.35$, a
60$''$ search radius corresponds to nearly 1 Mpc in diameter.  All extended
sources were visually inspected in the SDSS and FIRST images to give us
confidence that the radio emission is associated.  This does not mean that any
clearly associated emission out past 60$''$ was not included, but that any
associated emission out past 60$''$ was added to the total by hand after visual
inspection of the FIRST images.

In order to confirm the association of extended emission with the optical
counterpart it is first necessary to make sure that the extended emission is not
associated with another optical source in the field.  The majority of cases
where this takes place is when there is one core source within 4$''$ and a
second radio source within 60$''$, and where the second source is at the same
position as another galaxy in the SDSS image.  Therefore, any FIRST sources
found in the extended search with obvious optical counterparts in the SDSS
images were eliminated as possible extended source matches (e.g., SDSS
J094603.94+013923.6 is a BL AGN misclassified as a star in SDSS DR7 with a
resolved spiral galaxy to the north that is the likely the source of the
extended radio emission).  

The criteria used to confirm the association of the
extended emission to the central optical source are illustrated in Figure 1,
where the center of each image is the SDSS optical AGN position and the linear
scale is given below it.  In Figure 1, the sources a--e were all found
to have core emission in Paper I.  In Figure 1a (top left) we show the radio map
of SDSS J170013.70+400855.6, which has a core-source with a nearby ($\sim$ 35
kpc projected distance) knot of radio emission.  There is also a possible lobe
to the south-west that is below the flux limit of FIRST.  The association is
based on the fact that the second emission region is close to the host galaxy
and there is no optical source at or in the vicinity of the extended emission.
The sources that had only two emission regions turned out to be the most
numerous, and most were associated in this fashion.  In Figure 1b we show the
radio map of SDSS J122011.89+020342.2.  Here the association is based on the
physical connection of several emission knots in the eastern jet to the
core-source, and to a somewhat distant ($\sim$ 275 kpc projected distance) faint
lobe to the west.  In Figure 1c we show the radio map of SDSS
J132834.14-012917.6.  The association is based on the alignment of the very
distant lobes (both are $\sim$ 500 kpc projected distance) with the radio core
emission along with clear trails of radio emission back to the core.  There are
several variations of this type, such as those having small bending angles
(usually less than $\sim$ 15$^{\circ}$) between the distant lobes, as shown in
Figure 1d.  In Figure 1e we show the radio map of SDSS J091401.76+050750.6.  The
association is based on the distant southern lobe ($\sim$ 400 kpc projected
distance) having a hot spot and lobe emission structure that points back to the
core radio emission.  This object may in fact have an additional lobe source to
the north that is just outside of the image.  However, this was not added to the
total radio emission due to the fact that association at that distance is not
guaranteed without the other criteria being met. In this case, the exclusion of
this `could-be lobe' has very little effect on the conclusions due to the fact
that it is very dim and the added emission would have been only 4\% of the
total.  After visual inspection of all possible matches, we believe that there
are very few false positives (no more than 2 radio sources outside 4$''$ but inside
60$''$ that are not associated with the optical and radio core) in this search
when the objects with $<$ 4$''$ separation between radio core emission and
optical position are selected.

We find 51 (6\% of the original radio core emission sample and 0.6\% of the
total sample) AGN with extended emission that must be taken into account when
calculating the total AGN radio luminosity.  Of these 51, we find a large range
in the amount of extra emission that is picked up.  Some objects have a bright
core and one dimmer lobe ($\sim$ 10\%-50\% in added radio emission), but we also
find bright FR IIs that have total integrated fluxes in the 1000 mJy range
($\sim$ 100\%-600\% in added radio emission).  In order to characterize the
amount of flux added due to extended emission, we show in Figure 2 the fraction
of extended flux added with respect to the initial core emission.  About half of
the sources lie in the 0.01 -- 2 range, showing that nearly half of the sources
add only a small fraction and up to twice of the core flux to the total, while
the other half of the sample at least doubles the amount of flux added to the
core, and the brightest source adds nearly 70 times more emission when compared
to the core.

We performed a second search using the entire optical sample to find possible FR
II types in which radio emission is only seen from lobes but there is weak
(below the 1 mJy flux limit of FIRST) or no core radio emission.  The largest
group found in this search has just one single radio source that is within
60$''$.  After visual inspection, usually there is another optical source
matched to the extended radio source.  Even when there is no such alternative
optical identification it is not possible to claim an association since there is
no discernible jet to lobe connection or double lobe symmetry that would be
excellent indications of association.  Most of these were rejected outright.  We
do however find an additional 12 objects (not included in the 51 AGN discussed
above) that have significant flux inside the 60$''$ search radius, but no core
emission inside 4$''$, that can be positively associated with the optical
source.  All of these were visually inspected to ensure that the radio emission
was not associated with another optical source in the field and any clearly
associated emission at distances $>$ 60$''$ was taken into account and added
by hand to the total radio flux.  The criteria for establishing association for
these objects is the alignment of two sources of emission out past 4$''$ with
the optical source (having no detected radio core), as shown in Figure 1f for
SDSS J091519.56+563837.8.  We do note that any sources with radio lobes that
have significant bending angles would not satisfy our alignment criteria, and
some true associations may be excluded due to this effect.

\section{Results: Properties of the Extended Sources}
\indent

Table 1 lists the SDSS name of all 63 AGN with extended radio emission along
with their redshifts, projected physical extent and a `by eye' classification of
the radio morphology based on FIRST images.  There are 22 sources with no
previous radio identification in the NASA extragalactic database (NED) from
other radio surveys and they therefore have no radio catalogue source name in
Table 1.  The radio classification column makes use of the `giant radio galaxy'
(GRG) classification, where the total projected linear extent exceeds 750 kpc
\citep[ and references therein]{2009ApJ...695..156S}, and the hybrid morphology
radio sources (HYMORS) classification, where an FR II lobe is seen on one side
and an FR I jet is seen on the other side of the central source
\citep{2000A&A...363..507G}.  We also classify X-shaped radio sources
\citep[e.g.,][]{2003ApJ...594L.103G}, where a possible reorientation of the jets
has taken place to feed two individual sets of lobes, and the double-double
morphology (DDRG) where interruption of the jets can cause two distinct sets of
lobes to form throughout the lifetime of the AGN, where the first and older set
is at a larger distance than the second, younger pair
\citep[e.g.,][]{2000MNRAS.315..371S}.  
Table 2 summarizes the morphologies of the extended sources; the `indeterminate' 
designation is given to sources that were unresolved in the FIRST images.

We used the usual flux-luminosity relation with the same cosmology used by
\citet{Greene2007} from \citet{Spergel2003} (H$_{\rm 0}$ = 71 km s$^{-1}$
Mpc$^{-1}$, $\Omega_{\rm m}$ = 0.27, and $\Omega_{\rm \Lambda}$ = 0.73) to
calculate the total radio luminosity for each source.  From these new data, we
update the $\mathcal{R}$ values of the radio detected sample following Paper I,
including all associated extended emission.  Since the optical sample is the
same, those properties are unchanged.  We also break the sample into subsamples
as in Paper I.  The `detected sample' consists of all AGN with radio emission,
and contains the core-only sources and the extended sources.  The 63 extended
sources are all RL (i.e., all have $\mathcal{R}$ $>$ 10) with the exception of
one ($\mathcal{R} =$ 2.39) that is a face-on spiral galaxy (SDSS
J220233.84-073225.0) with a modest L$_{\rm H\alpha}$ = 10$^{42.01}$ ergs
s$^{-1}$, but a very low flux radio core ($F_{\rm int} =$ 2.36 mJy) with an even
fainter `lobe' (F$_{\rm int} =$ 0.97 mJy) offset by $10\farcs5$.  We find that
of the 793 remaining core-only sources, 383 are RL and 410 are RQ, based on
$\mathcal{R}$.  The `flux limited sample' is explicitly defined in Paper I and
has 5485 total objects, using an upper limit of 1 mJy for all AGN without radio
detections as an optical flux cutoff.  For the flux limited sample we find that
4.9\% (270/5485) of the AGN are RL compared to the 4.7\% (259/5485) found in
Paper I when extended emission was not taken into account.

In order to determine how the extended sources differ from the rest of the
sample we first compare the optical properties of the different subsets, namely
the core-only sources, the extended sources and the total flux limited sample. 
In Figure 3 we show the FWHM distributions of the broad component of the
H$\alpha$ line (FWHM$_{\rm H\alpha}$) for the extended, core-only, and non-radio
detected sources in the flux limited sample, where all are normalized by the
number in each group.  The extended source distribution has an average FWHM of
5010 km s$^{-1}$ and the core-only sources have an average FWHM of 3550 km
s$^{-1}$.  The peak of the distribution for the extended sources is at $\sim$
4500 km s$^{-1}$ and is shifted to higher FWHM values by about 2000 km s$^{-1}$
when compared to the core-only sources, that peak at $\sim$ 2500 km s$^{-1}$. 
Both histograms have long tails that fall off at about the same rate toward
higher FWHMs.  A K-S test between the core-only source distribution and the
extended source distribution yields a probability value of 1.2 $\times 10^{-5}$
and a maximum difference of 0.35, showing that it is extremely likely these two
distributions are from different parent populations.  A K-S test between the
core-only sources and the non-radio detected sources yields a probability value
of 0.077 and a maximum difference of 0.05, meaning that the two have similar
enough cumulative distribution functions that they may well be from the same
parent population.  Visual inspection of the optical spectra shows that many of
the extended AGN have characteristically wide H$\alpha$ profiles.  This is
consistent with the claim that most BLRGs have intrinsically large M$_{\rm BH}$
\citep{2003astro.ph.12417L, 2003MNRAS.340.1095D, 2005ApJ...625..716C}.  This
result is of course favored when calculating M$_{\rm BH}$ based on single epoch
M$_{\rm BH}$ relations, where M$_{\rm BH} \propto$ FWHM$^{2}$, but a large
M$_{\rm BH}$ determination is not always guaranteed since this relation also
depends on the optical luminosity of the central source (M$_{\rm BH}$ $\propto$
FWHM$^{2}$L$^{0.5}$).

The next optical property we compare between the radio types is the H$\alpha$
luminosity ($\propto$ L$_{\rm 5100}$).  The histogram in Figure 4 shows the
normalized distributions for the extended, core-only, and non-radio detected
sources in the flux limited sample.  We find that the extended source
distribution is shifted to higher L$_{\rm H\alpha}$ by about 0.5 dex when
compared to the core-only distribution, but overall the full distributions have
similar peak values and show significant overlap.  More precisely, the extended
sources have an average L$_{\rm H\alpha} = 10^{42.7}$ ergs s$^{-1}$ with a
standard deviation of 0.60 dex, and the core-only sources have an average
L$_{\rm H\alpha} = 10^{42.3}$ ergs s$^{-1}$ with a standard deviation of 0.68
dex.  A K-S test comparing the extended sources and the core-only sources yields
a probability value of 1.4 $\times 10^{-4}$ and a maximum difference of 0.32
indicating that these two distributions may well be from different parent
populations.  In the context of M$_{\rm BH}$ determinations, somewhat similar
L$_{\rm H\alpha}$ distributions but systematically higher FWHM distributions
should give larger M$_{\rm BH}$ estimates for the extended AGN when compared to
the core-only sources.  This turns out not to always be the case, since our
extended sample has 36 sources with M$_{\rm BH} < 10^{8}$ M$_{\odot}$ and 27
sources have M$_{\rm BH} > 10^{8}$ M$_{\odot}$.

The normalized distribution of 1.4 GHz radio luminosity (L$_{\rm 1.4 GHz}$) is
shown in Figure 5 for the flux limited sample.  The peak of the extended sources
is shifted to higher luminosities by a factor of 100 when compared to the
core-only sources.  This is not surprising given the high luminosities of FR II
lobes.  Looking at the region of overlap we find that there are few sources in
these normalized distributions in the 10$^{24.5}$ Watts Hz$^{-1}$ region, where
the deficit of sources is at the FR I/FR II transition luminosity originally
found by \citet{1974MNRAS.167P..31F}; see also \citet{Kawakatu2009}.  This is
important for the log($\mathcal{R}$) histogram shown in Figure 6.  In the top
plot we show the core-only and extended source histograms normalized by the
number in each group.  The normalization is useful since there are many fewer
extended radio sources, making this trend in the unnormalized histogram not as
obvious.  Here we find what looks like two separate populations, or an apparent
bimodality, in that the extended sources peak at log($\mathcal{R}$) $\approx$
2.5 whereas the core-only sources peak at about log($\mathcal{R}$) $\approx$
0.75.  This can be explained by the fact that most of the extended sources have
much higher radio luminosities compared to the core-only sources, but not much
higher H$\alpha$ luminosities, causing the shift of extended sources to higher
log($\mathcal{R}$) values.  This produces a bimodal distribution where the upper
mode is comprised of RL objects ($\mathcal{R} >$ 10) populated by only the
extended sources whose distribution drops below log($\mathcal{R}$) = 1 only for
the one RQ source mentioned above.  The core-only distribution, however, goes
well above and below the log($\mathcal{R}$) = 1 RL/RQ dividing line.  

The bottom plot in Figure 6 shows the histogram of two different populations
from the detected sample (core-only and extended sources) based on a radio
luminosity dividing line.  The value of the radio luminosity dividing line was
found by adjusting the break luminosity value until the lower histograms best
matched the original core-only versus extended source histograms shown in the
upper plot (based on K-S statistics given below), and was found to be
10$^{24.4}$ Watts Hz$^{-1}$.  It is interesting that the two sets of histograms
are most similar when the break radio luminosity is nearly equal to that of the
FR I/FR II transition luminosity.  It is clear that the AGN with L$_{\rm r} <
10^{24.4}$ Watts Hz$^{-1}$ have a log($\mathcal{R}$) distribution nearly
identical to the core-only sources and the AGN with L$_{\rm r} > 10^{24.4}$
Watts Hz$^{-1}$ have a log($\mathcal{R}$) distribution nearly identical to the
extended sources. A K-S test comparing the extended sources in the top plot and
the FR II-like distribution in the bottom plot yield a probability value of 1.0
and a maximum difference of 0.04, showing that it is extremely unlikely that
these two are from different parent populations.  This is also found for the
core-only and FR I-like distributions, which have a K-S probability value of
0.82 and a maximum difference of 0.03.

Therefore, in order to move away from the `by eye' morphological classification
schemes used to describe individual sources, we can in general use our extended
sources as a proxy for the classic FR II objects, and the core-only sources as a
proxy for the FR I sources based on a break radio luminosity that is consistent
with the previous FR I/FR II dividing line.  From this plot we also find that a
log($\mathcal{R}$) value of $\approx$ 1.75 is well suited to separate the FR Is
from the FR IIs.  The peak values of the $\mathcal{R}$ histogram are consistent
with the bimodal distributions found by the previous studies of
\citet{Ivezic2004} and \citet{Cirasuolo2004} who find peaks at
log($\mathcal{R}$) $<<$ 1, and log($\mathcal{R}) \approx$ 2-3 using only the
radio core sources out to higher redshifts, which may possess complex (extended)
structure, but which would be unresolvable at the higher redshifts probed in
these samples.  Here we show that our two populations basically consist of the
lower radio power FR Is (which could be young, unresolved or well aligned with
the line-of-sight jets and/or lobes) and the higher radio power FR IIs, and that
the $\mathcal{R}$ bimodality seen here is likely a manifestation of the FR I/FR
II break originally found by \citet{1974MNRAS.167P..31F}.

We updated the radio luminosities of the AGN in our sample to determine
the effects of the extended radio emission on our previous results in Paper I.
As might be expected, the additional flux in a small fraction ($\sim$8\%) of the
radio-detected sample had little effect on the overall trends that we found
between radio loudness and Eddington ratio and/or black-hole mass (see
\citet{Rafter2010} for the updated plots).

\section{Conclusions}
We have taken the SDSS BL AGN sample from \citet{Greene2007} and performed a
search for extended associated radio emission using the VLA's FIRST survey.  We
find that 846 of the objects (10\%) have core emission and 63 (0.8\%) have
extended emission that must be taken into account when calculating the total
radio luminosity and radio-loudness.  We compare these results to \citet{Rafter2009} and find that the trends in radio-loudness with other physical properties are largely unchanged, which is unsurprising as the detected sample was only modestly enlarged overall.  The RLF as a function of L$_{\rm bol}$/L$_{\rm Edd}$ and M$_{\rm BH}$ are essentially the same, and we still find a modest trend of decreasing RLF with increasing log(L$_{\rm bol}$/L$_{\rm Edd}$), along with an increase of the RLF as M$_{\rm BH}$
increases above $\sim 2 \times 10^{8}$ M$_{\odot}$.  We do note that about half
of the extended RL AGN do {\it not} have the most massive BHs (M$_{\rm BH} >
10^{8}$ M$_{\odot}$), indicating that even extreme radio-loudness is not based
solely on M$_{\rm BH}$, but must also be closely tied to other fundamental
parameters such as black hole spin or accretion mode, although our data do not allow us
to draw conclusions as to which, if either, of those theoretical paradigms for radio power
is more likely to be correct. 

With extended emission taken into account, we find evidence
for a distinct population of RL AGN comprised of the extended sources that is
separate from the RL and RQ core-only sources.  We find that most of the extended AGN
in this low redshift sample are FR IIs based on radio morphology and luminosity,
using the same FR I/FR II break luminosity defined by \citet{1974MNRAS.167P..31F}.  We find a bimodal distribution in the $\mathcal{R}$ parameter, but at a value above the classic RL/RQ dividing line and propose that this is a manifestation of the FR I/FR II break.  In the previous high redshift studies mentioned above, where only the `core' radio
emission is used, the bimodality in $\mathcal{R}$ may again be a manifestation of the FR I/FR II transition, although what is considered to be core emission may in fact include jets and/or lobes (or relatively young sources) whose true radio structure is unresolved due to their extreme distances.  We do note that for the sources with just two components, where one is the core source and the other is not at a large angular distance, the morphology is not easy to determine based on the resolution of the FIRST survey.

The distributions of optical luminosity for the H$\alpha$ emission line (Figure 4) are more
similar for both the core-only and extended sources, as well as the total
sample, than are the distributions of radio luminosities (Figure 5).  This difference gives
rise to the radio dichotomy seen when evaluating radio-loudness based on the
$\mathcal{R}$ parameter in this sample.  From our sample we propose that a more
interesting dividing line is at a log($\mathcal{R}$) value of $\sim$ 1.75
instead of the classical log($\mathcal{R}) = 1$.  This higher break value for
log($\mathcal{R}$) separates local broad-line AGN into two distinct populations
of undetected/core-only radio sources and extended radio sources in the FIRST
survey. 

The claims of bimodality between RL and RQ AGN have usually been based on the somewhat arbitrary log($\mathcal{R}$) $=$ 1 criteria, and were heavily debated based on sample selection and inclusion/exclusion criteria (see Section 1 and references therein). While the dichotomy between RL and RQ AGN is called into question in the studies of \citet{2007ApJ...654...99W} and \citet{Rafter2009}, the study by \citet{Sikora2007} does find evidence for this dichotomy and further, postulates physical conditions that may be responsible for its existence.  In this work we can clearly reproduce a dichotomy between the core-only sources, comprised of RQ and weak/unresolved FR I type AGN, against the most powerful FR I and FR II type AGN in our sample.  Such a distinction may have a more physical and theoretically compelling basis as opposed to being a distinction between RL and RQ that is influenced strongly by observational constraints.  As shown in Figure 6, all the AGN in the upper population have extended emission and resolved complex morphologies whose BH/accretion disk system may have different or extreme physical properties when compared to the much more numerous undetected and core-only sources.  The lower population (core-only) sources may then be made up of two types. The first type, where the radio emission originates from coronal emission on subparsec scales, as discussed in \citet{2008MNRAS.390..847L}, could constitute the bulk of objects with log($\mathcal{R}$) $<$ 1.  The second type could contain either unresolved young jets which emit on a scale of a few pc and/or weak jets with intrinsically weak radio emission and low kinetic jet power.  This second type of object would still fit into our lower population while having 1.75 $>$ log($\mathcal{R}$) $>$ 1, thereby satisfying the classic paradigm that FR Is tend to be RL based on log($\mathcal{R}$) $>$ 1.  Once the threshold of log($\mathcal{R}$) $=$ 1.75 is crossed in Figure 6, there is a clear transition to the most radio powerful AGN, with strong jets and bright extended emission; these are plausibly a result of some difference in accretion mode, accretion rate, or BH spin in the central engine, whereby the efficiency of jet launching is greatly enhanced.

\acknowledgments
We thank Jenny Greene and Luis Ho for providing their data set and for helpful
advice.  PJW was supported in part by a subcontract to GSU from NSF grant
AST05-07529 to the University of Washington.  Funding for the SDSS Archive has
been provided by the Alfred P. Sloan Foundation, the Participating Institutions,
the National Aeronautics and Space Administration, the National Science
Foundation, the U.S. Department of Energy, the Japanese Monbukagakusho, and the
Max Planck Society. The SDSS Web site is http://www.sdss.org/.  The SDSS is
managed by the Astrophysical Research Consortium (ARC) for the Participating
Institutions. The Participating Institutions are The University of Chicago,
Fermilab, the Institute for Advanced Study, the Japan Participation Group, The
Johns Hopkins University, Los Alamos National Laboratory,the
Max-Planck-Institute for Astronomy (MPIA), the Max-Planck-Institute for
Astrophysics (MPA), New Mexico State University, Princeton University, the
United States Naval Observatory, and the University of Washington.  The FIRST
Survey is supported by grants from the National Science Foundation, NATO, the
National Geographic Society, Sun Microsystems, and Columbia University.

\begin{deluxetable}{llcccl}
\rotate
\tablecolumns{6}
\tabletypesize{\footnotesize}
\tablewidth{0pc}
\tablecaption{The Matched SDSS and FIRST Sample\label{table}}
\tablehead{
	\colhead{SDSS Name} & 
	\colhead{Radio Catalogue} & 
	\colhead{Redshift} &
	\colhead{Total Integrated} & 
	\colhead{Projected Physical} & 
	\colhead{Radio Classification} \\
	\colhead{} & 
	\colhead{and Source Name} & 
	\colhead{} & 
	\colhead{Flux (mJy)} & 
	\colhead{Size (Mpc)} &
	\colhead{} \\
\colhead{(1)} & \colhead{(2)} & \colhead{(3)} & \colhead{(4)} &
\colhead{(5)} & \colhead{(6)}
}

\startdata
J005550.75-101905.6 & FBQS J0055-1019		      & 0.3091 &   48.58 & 0.94 & GRG/FR II		       \\  
J013352.65+011345.3 & 87GB 013118.8+005811	      & 0.3081 &   67.15 & 0.62 & FR II			       \\
J072406.79+380348.6 &				      & 0.2413 &  203.46 & 0.58 & FR II			       \\
J074906.50+451033.9 & B3 0745+453,GB6 J0749+4510      & 0.1921 &  117.74 & 0.12 & Core + weak jet(5\%)	       \\
J075244.19+455657.3 & B30749+460A,6CB074906.2+460422  & 0.0518 &  238.41 & 0.14 & FR I			       \\
J075643.09+310248.7 &				      & 0.2715 &   22.59 & 0.14 & Classic Triple FR II	       \\
J080129.57+462622.8 &				      & 0.3159 &   13.37 & 0.26 & Classic Triple FR II	       \\
J082133.60+470237.2 & 3C 197.1, *B3 0818+472A	      & 0.1280 & 1711.27 & 0.06 & Bright FR II	   	       \\
J082355.36+244830.4 &				      & 0.2339 &    2.32 & 0.06 & Faint FR II		       \\
J084600.36+070424.6 & 87GB 084319.4+071534	      & 0.3421 &  241.53 & 0.85 & GRG/FR II		       \\
J085348.18+065447.1 & PMN J0853+0654		      & 0.2232 &  769.90 & 0.08 & Core + 1 Bright lobe	       \\
J085627.91+360315.6 &				      & 0.3449 &   29.96 & 0.20 & FR II			       \\
J091133.85+442250.1 & B3 0908+445,GB6 J0911+4422      & 0.2976 &  433.23 & 0.15 & FR I			       \\
J091401.76+050750.6 & 4C +05.38			      & 0.3014 &  328.72 & 0.46 & Large FR II lobe in SW	\\
J091519.55+563837.8 &				      & 0.2631 &   19.98 & 0.56 & FR II	    		       \\
J092308.16+561455.3 &				      & 0.2493 &  143.01 & 0.23 & FR II			       \\
J092837.97+602521.0 & 8C 0924+606		      & 0.2955 &  278.21 & 0.25 & FR II			       \\
J093200.08+553347.4 & 6C B092828.4+554656 	      & 0.2657 &   73.43 & 0.94 & GRG/FR II			\\
J094144.82+575123.6 & GB6 J0941+5751		      & 0.1585 &   90.43 & 0.11	& FR II			       \\
J094745.14+072520.5 & 3C 227, PKS 0945+07 	      & 0.0858 & 3117.09 & 0.40 & FR II			       \\
J095456.89+092955.8 & 4C +09.35, PKS 0952+097 	      & 0.2984 &  440.66 & 0.17 & FR II			       \\
J100726.10+124856.2 & 4C +13.41, PKS 1004+13 	      & 0.2406 &  959.14 & 0.52 & FR II			       \\
J100819.11+372903.4 &				      & 0.0522 &    2.27 & 0.02 & 2nd source in host galaxy    \\
J103143.51+522535.1 & 4C +52.22, GB6 J1031+5225	      & 0.1662 &  904.01 & 0.13 & FR II			       \\
J103458.35+055231.8 &				      & 0.3002 &   28.72 & 0.15 & one lobe SW		       \\
J105220.30+454322.2 &				      & 0.2406 &  112.05 & 0.26 & FR I (asymmetric)  	       \\
J105500.33+520200.9 & 6C B105202.4+521804	      & 0.1874 &  461.07 & 0.21 & FR II			       \\
J105632.01+430055.9 &				      & 0.3177 &   19.37 & 0.22 & FR II			       \\
J110845.48+020240.8 & PKS 1106+023		      & 0.1574 &  784.08 & 0.08 & Core + possible lobe	       \\
J111432.79+105034.7 &				      & 0.1931 &  780.25 & 0.23 & DDRG/FR II			\\
J113021.40+005823.0 & 4C +01.30, PKS 1127+012	      & 0.1323 &  566.72 & 0.16 & X-shaped (0.26Mpc) 	       \\
J114004.35-010527.4 & [WB92] 1137-0048 		      & 0.3470 &   34.17 & 1.12 & GRG/HYMORS		       \\
J114047.90+462204.8 & 87GB 113808.0+463858 	      & 0.1149 &   91.99 & 0.06 & core + bent jet	       \\
J114958.70+411209.4 & 6C B114721.6+412848 	      & 0.2497 &  118.46 & 0.33 & FR I			       \\
J115409.27+023815.0 & 87GB 115136.0+025423	      & 0.2106 &   64.12 & 0.26 & FR I			       \\
J115420.72+452329.4 &				      & 0.1912 &  964.77 & 0.29 & FR II			       \\
J120612.67+490226.2 &				      & 0.1194 &    6.30 & 0.09 & Possible core-only source    \\
J122011.89+020342.2 & PKS 1217+02		      & 0.2404 &  482.78 & 0.57 & FR I (asymmetric and bent)    \\
J123807.77+532555.9 & 87GB123550.3+534219	      & 0.3475 &   61.60 & 1.02 & GRG/FR II		       \\
J123915.39+531414.6 & 6C B123659.8+533024 	      & 0.2013 &   23.11 & 0.26 & FR II w/ faint core	       \\
J130359.47+033932.1 & 4C +03.26			      & 0.1837 &  210.85 & 0.45 & FR II			       \\
J131827.00+620036.2 & 87GB131634.0+621623,8C 1316+622 & 0.3075 &  133.41 & 0.38 & FR II			       \\
J132404.20+433407.1 &				      & 0.3377 &  239.62 & 1.10 & GRG/FR II		       \\
J132834.14-012917.6 &				      & 0.1514 &  158.85 & 0.98 & GRG/FR II		       \\
J133253.27+020045.6 & 3C 287.1 			      & 0.2158 & 1759.16 & 0.57 & FR II			       \\
J133437.48+563147.9 & 87GB133243.4+564710	      & 0.3428 &  164.18 & 0.24 & FR I			       \\
J134545.35+533252.3 & 87GB 134352.4+534755 	      & 0.1354 &  278.19 & 0.13 & FR II		   	       \\
J134617.54+622045.4 & 6C B134441.6+623604 	      & 0.1164 &  142.99 & 0.15 & FR I (bent)		       \\
J141613.36+021907.8 &				      & 0.1582 &  107.70 & 0.67 & DDRG/FR II		       \\
J144302.76+520137.2 & 3C 303			      & 0.1412 & 2119.27 & 0.12 & FR II			       \\
J151640.22+001501.8 & GB6 J1516+0015, 4C +00.56       & 0.0524 & 1090.21 & 0.28 & FR II			       \\
J151913.35+362343.4 & 6C B151717.1+363448	      & 0.2857 &  207.25 & 0.58 & HYMORS candidate	       \\
J152942.20+350851.2 & 7C 1527+3519,6C B152745.2+35192 & 0.2873 &  109.27 & 0.08 & Bright core + 1 lobe	       \\
J155206.58-005339.3 &   			      & 0.2977 &  105.67 & 0.12 & FR I (partially resolved)     \\
J163856.53+433512.5 & B3 1637+436A,6CB163723.1+434051 & 0.3390 &  133.04 & 0.45 & FR II			       \\
J164442.53+261913.2 &				      & 0.1442 &  110.36 & 0.06 & Unresolved core structure    \\
J170013.70+400855.6 &				      & 0.0941 &   20.68 & 0.07 & faint lobe SW, FR II?	       \\
J170425.11+333145.9 &				      & 0.2902 &   36.07 & 0.37 & FR II			       \\
J171322.58+325628.0 & FBQS J171322.6+325628	      & 0.1013 &   44.80 & 0.15 & faint FR I		       \\
J220233.84-073225.0 &				      & 0.0594 &    3.33 & 0.02 & RQ,2nd source in host galaxy \\
J230545.66-003608.6 & 4C -01.59, PKS 2303-008	      & 0.2689 &  517.76 & 0.15 & FR II			       \\
J233313.16+004911.8 & PKS 2330+005		      & 0.1700 &  317.86 & 0.17 & FR I			       \\
J235156.12-010913.3 & 4C -01.61, PKS 2349-01 	      & 0.1740 & 1460.41 & 0.09 & FR II			       \\

\enddata
\tablecomments{
Col.(1): SDSS Name;
Col.(2): Radio Catalogue and Source Name: taken from the NASA Extragalactic Database (NED);
Col.(3): Redshift: taken from SDSS spectra;
Col.(4): Total Integrated Radio Flux at 1.4 GHz (mJy);
Col.(5): Projected Physical Size: these approximate values are calculated using the FIRST radio maps (Mpc);
Col.(6): Radio Classification: Radio Quiet (RQ), Fanaroff \& Riley class 1 \& 2 (FR I, FR II respectively), Giant Radio Galaxy (GRG), HYbrid MOrphology Radio Source (HYMORS), Double-Double Radio Galaxy (DDRG), X-shaped (having radio emission that resembles an `x' pattern, where there are two sets of symmetric emission regions at $\sim$ 90$^{\circ}$ to each other).}
\end{deluxetable}

\begin{deluxetable}{lr}
\tabletypesize{\footnotesize}
\tablewidth{0pc}
\tablecaption{Summary of Radio Morphologies\label{table2}}
\tablehead{
	\colhead{Radio Morphology} & 
	\colhead{Number} \\ } 

\startdata

FR I            &       10      \\
FR II           &       25      \\
FR II/GRG       &        6      \\
DDRG            &        2      \\
X$-$shaped      &        1      \\
HYMORS          &        2      \\
Indeterminate   &       17      \\ 

\enddata
\end{deluxetable}

\begin{figure}[b]
\includegraphics[angle=0, scale=.75]{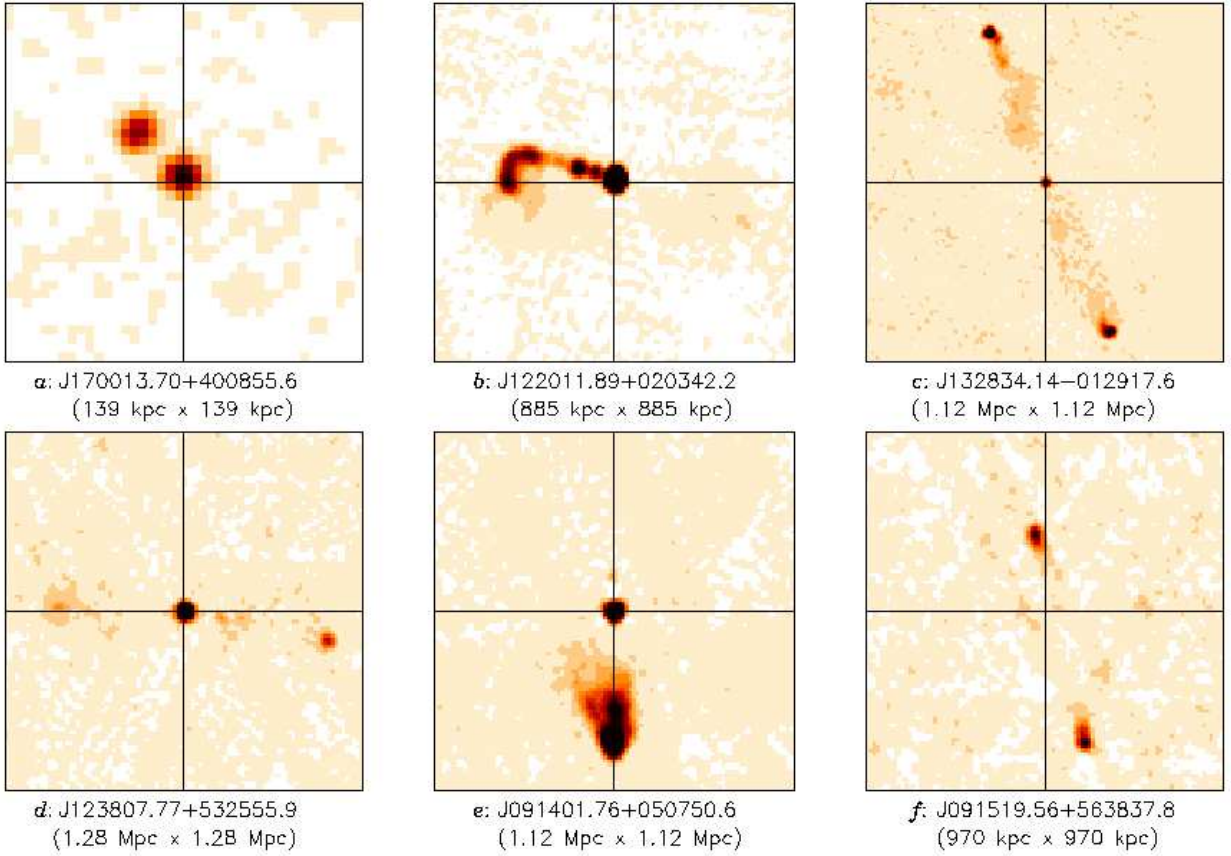}
\caption[]{Images of extended FIRST sources with the optical source at the center of each frame, north is up and east is to the left.  The SDSS name and projected physical size is given beneath each image.  \textbf{a:} The $1\farcm2 \times 1\farcm2$ image shows a double source with one component on the optical core and one offset from it. \textbf{b:} The $3\farcm0 \times 3\farcm0$ image has a strong jet and lobe to the east and a weaker lobe to the west. \textbf{c:} The $6\farcm0 \times 6\farcm0$ image shows a giant FR II where the lobe to the NE is aligned with the radio core and lobe to the SW. \textbf{d:} The $3\farcm0 \times 3\farcm0$ image shows a distant lobe to the east, a radio core, and a lobe to the west slightly misaligned. \textbf{e:} The $3\farcm0 \times 3\farcm0$ image shows a giant radio lobe with multiple sources to the south that all point back to the radio core.  Not shown in this image is a more distant and slightly misaligned source to the north that may be an associated lobe, but with very low flux.  \textbf{f:} The $3\farcm0 \times 3\farcm0$ image shows two lobes that are roughly aligned with the optical center, but with no detected radio core.}
\label{fig1}
\end{figure}

\begin{figure}[b]
\includegraphics[angle=90, scale=0.6]{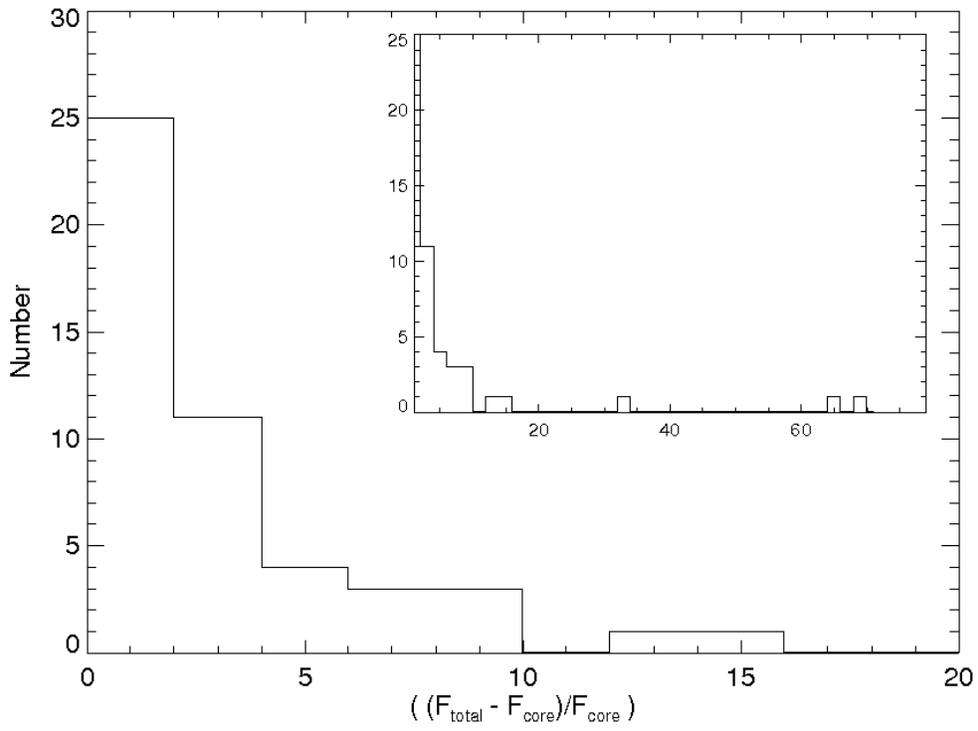}
\caption[]{Histogram characterizing the amount of extended radio flux detected for the 51 extended sources with core emission.  Nearly half (25) of the sources add only a fraction and up to two times the core flux.  The inset shows that the largest increase in total flux due to extended emission is nearly 70 times the core flux.}
\label{fig2}
\end{figure}

\begin{figure}[b]
\includegraphics[angle=90, scale=0.6]{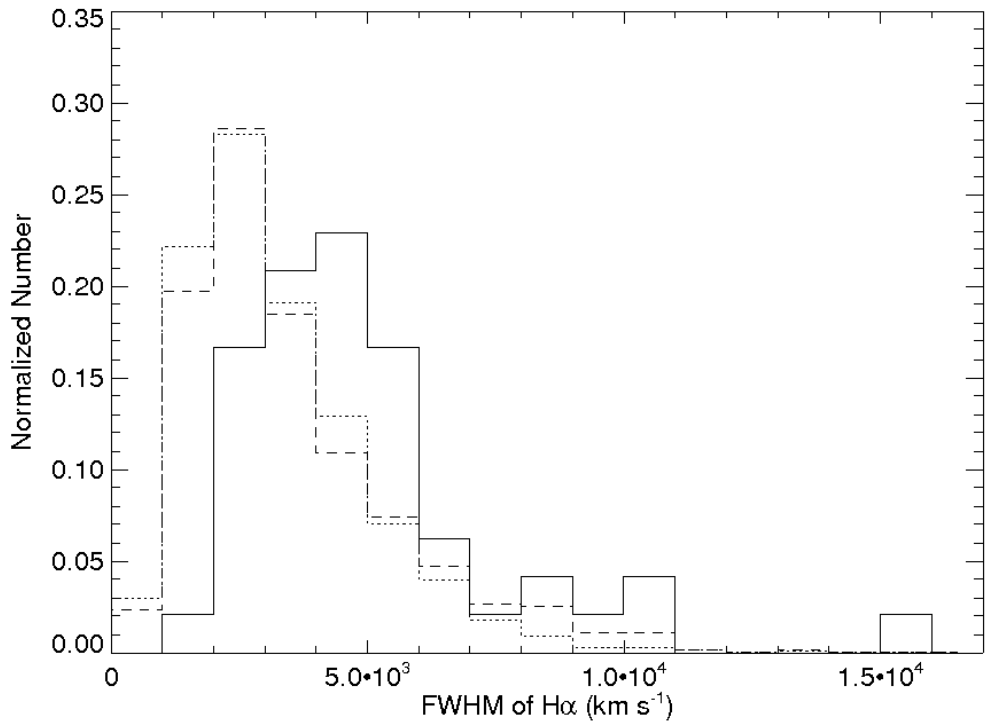}
\caption[]{FWHM$_{\rm H\alpha}$ histogram: the solid line is for the extended sources, the dashed line is for the core-only sources, and the dotted line is for the non-radio detected sources in the flux limited sample.  The extended source distribution is shifted to higher values by $\sim$ 2000 km s$^{-1}$ compared to the core-only and flux limited samples.}
\label{fig3}
\end{figure}

\begin{figure}[b]
\includegraphics[angle=90, scale=0.6]{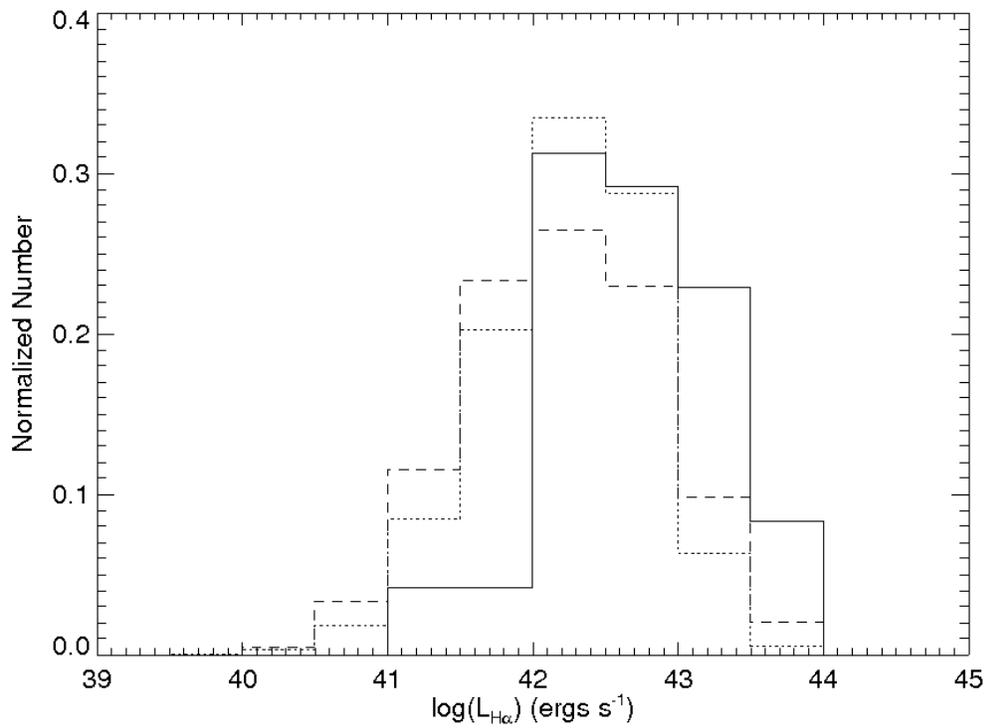}
\caption[]{L$_{\rm H\alpha}$ histogram: the solid line is for the extended sources, the dashed line is for the core-only sources, and the dotted line is for the non-radio detected sources in the flux limited sample.  The distributions have similar shapes, peak values, and show significant overlap at luminosities greater than 10$^{42}$ ergs s$^{-1}$.}
\label{fig4}
\end{figure}

\begin{figure}[b]
\includegraphics[angle=90, scale=0.6]{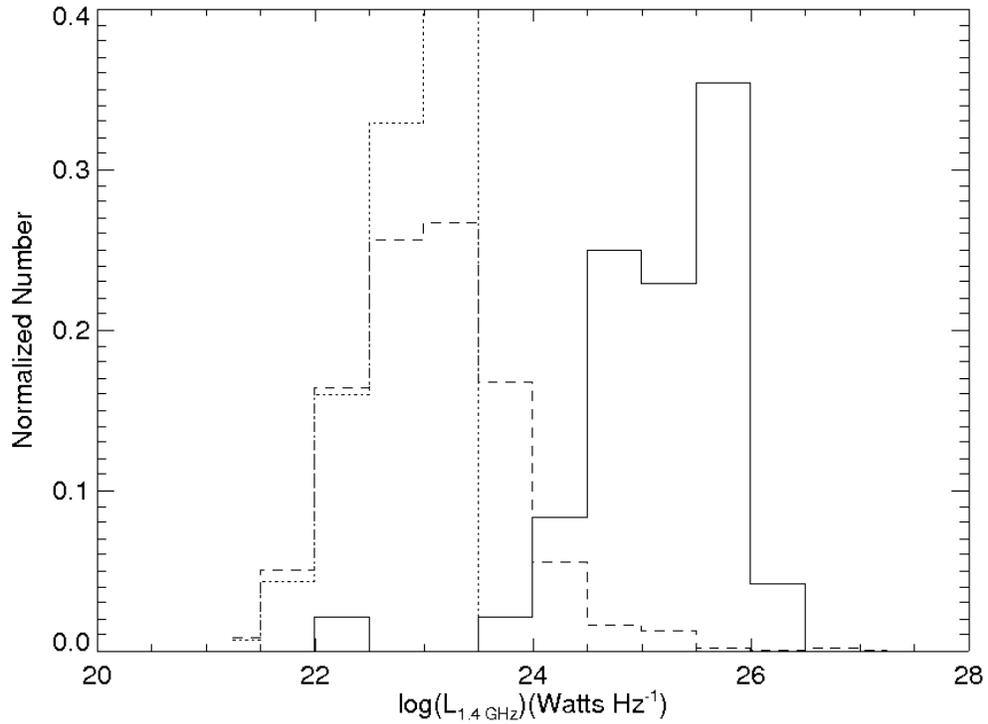}
\caption[]{L$_{\rm 1.4 ~GHz}$ histogram: the solid line is for the extended sources, the dashed line is for the core-only sources, and the dotted line is for the full optical sample.  The sharp drop off of the dotted line at 10$^{23.5}$ Watts Hz$^{-1}$ is due to normalization and does not actually go to zero.  The relative lack of sources at 10$^{24.5}$ Watts Hz$^{-1}$ is a manifestation the FR I/FR II dividing line.}
\label{fig5}
\end{figure}

\begin{figure}[b]
\includegraphics[angle=90, scale=0.6]{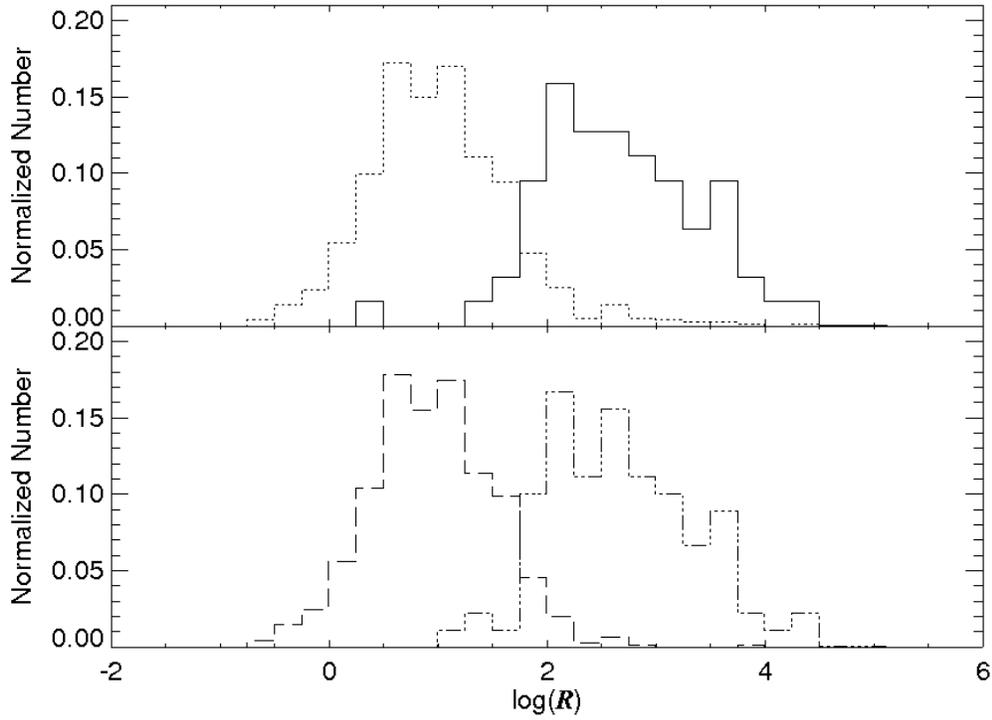}
\caption[]{log($\mathcal{R}$) histogram: Top, the solid line is for the extended sources and the dotted line is for the core-only sources.  Bottom,  for the combined sample of core-only and extended sources the dashed line is for objects with L$_{1.4 GHz} < 10^{24.4}$ Watts Hz$^{-1}$ and represents an FR I-like population and the dot-dashed line is for objects with L$_{1.4 GHz} > 10^{24.4}$ Watts Hz$^{-1}$ and represents an FR II-like population.}
\label{fig6}
\end{figure}

\clearpage


\begin{thebibliography}

\bibitem[Antonucci(1993)]{1993ARA&A..31..473A} Antonucci, R.\ 1993, \araa, 31, 473 

\bibitem[Becker et al.(1995)]{Becker1995} Becker, R.~H., White, 
R.~L., \& Helfand, D.~J.\ 1995, \apj, 450, 559

\bibitem[Bentz et al.(2009)]{2009ApJ...697..160B} Bentz, M.~C., Peterson, 
B.~M., Netzer, H., Pogge, R.~W., \& Vestergaard, M.\ 2009, \apj, 697, 160

\bibitem[Best et al.(2005)]{Best2005} Best, P.~N., Kauffmann, 
G., Heckman, T.~M., Brinchmann, J., Charlot, S., Ivezi{\'c}, {\v Z}., 
\& White, S.~D.~M.\ 2005, \mnras, 362, 25

\bibitem[Blandford 
\& Znajek(1977)]{1977MNRAS.179..433B} Blandford, R.~D., \& Znajek, R.~L.\ 1977, \mnras, 179, 433

\bibitem[Chiaberge et al.(2005)]{2005ApJ...625..716C} Chiaberge, M., 
Capetti, A., \& Macchetto, F.~D.\ 2005, \apj, 625, 716

\bibitem[Cirasuolo et al.(2004)]{Cirasuolo2004} Cirasuolo, M., 
Celotti, A., Magliocchetti, M., 
\& Danese, L.\ 2004, in AGN Physics with the Sloan Digital Sky Survey, ASP Conf.\ Ser.\ Vol.\ 311, ed.\
G.\ T.\ Richards, P.\ B.\ Hall (San Franscisco: ASP), p.\ 343

\bibitem[Dunlop et al.(2003)]{2003MNRAS.340.1095D} Dunlop, J.~S., McLure, 
R.~J., Kukula, M.~J., Baum, S.~A., O'Dea, C.~P., 
\& Hughes, D.~H.\ 2003, \mnras, 340, 1095

\bibitem[Elvis et al.(1994)]{1994ApJS...95....1E} Elvis, M., et al.\ 1994, 
\apjs, 95, 1

\bibitem[Elvis et al.(2002)]{2002ApJ...565L..75E} Elvis, M., Risaliti, G., 
\& Zamorani, G.\ 2002, \apjl, 565, L75

\bibitem[Fanaroff 
\& Riley(1974)]{1974MNRAS.167P..31F} Fanaroff, B.~L., \& Riley, J.~M.\ 1974, \mnras, 167, 31P 

\bibitem[Gopal-Krishna 
\& Wiita(2000)]{2000A&A...363..507G} Gopal-Krishna, \& Wiita, P.~J.\ 2000, \aap, 363, 507

\bibitem[Gopal-Krishna et al.(2003)]{2003ApJ...594L.103G} Gopal-Krishna, 
Biermann, P.~L., \& Wiita, P.~J.\ 2003, \apjl, 594, L103

\bibitem[Gopal-Krishna et al.(2008)]{2008ApJ...680L..13G} Gopal-Krishna, 
Mangalam, A., \& Wiita, P.~J.\ 2008, \apjl, 680, L13

\bibitem[Greene 
\& Ho(2004)]{Greene2004} Greene, J.~E., \& Ho, L.~C.\ 2004, \apj, 610, 722

\bibitem[Greene 
\& Ho(2007)]{Greene2007} Greene, J.~E., \& Ho, L.~C.\ 2007, \apj, 667, 131

\bibitem[Ho(2002)]{2002ApJ...564..120H} Ho, L.~C.\ 2002, \apj, 564, 120

\bibitem[Ivezi{\'c} et al.(2002)]{Ivezic2002} Ivezi{\'c}, {\v Z}., 
et al.\ 2002, \aj, 124, 2364

\bibitem[Ivezi{\'c} et al.(2004)]{Ivezic2004} Ivezi{\'c}, Z., et 
al.\ 2004, in AGN Physics with the Sloan Digital Sky Survey, ASP Conf.\ Ser.\ Vol. 311, ed.\ G.\ T.\ Richards, P.\ B.\ Hall (San Francisco: ASP), p.\ 347

\bibitem[Kaspi et al.(2000)]{Kaspi2000} Kaspi, S., Smith, P.~S., 
Netzer, H., Maoz, D., Jannuzi, B.~T., \& Giveon, U.\ 2000, \apj, 533, 631

\bibitem[Kaspi et al.(2005)]{Kaspi2005} Kaspi, S., Maoz, D., 
Netzer, H., Peterson, B.~M., Vestergaard, M., 
\& Jannuzi, B.~T.\ 2005, \apj, 629, 61

\bibitem[Kawakatu et al.(2009)]{Kawakatu2009} Kawakatu, N., Kino, 
M., \& Nagai, H.\ 2009, \apjl, 697, L173

\bibitem[Kellermann et al.(1989)]{Kellermann1989} Kellermann, K.~I., 
Sramek, R., Schmidt, M., Shaffer, D.~B., \& Green, R.\ 1989, \aj, 98, 1195

\bibitem[Kewley et al.(2006)]{2006MNRAS.372..961K} Kewley, L.~J., Groves, 
B., Kauffmann, G., \& Heckman, T.\ 2006, \mnras, 372, 961

\bibitem[Laor(2000)]{Laor2000} Laor, A.\ 2000, \apjl, 543, L111

\bibitem[Laor(2003)]{2003astro.ph.12417L} Laor, A.\ 2003, ArXiv 
Astrophysics e-prints, arXiv:astro-ph/0312417

\bibitem[Laor 
\& Behar(2008)]{2008MNRAS.390..847L} Laor, A., \& Behar, E.\ 2008, \mnras, 390, 847

\bibitem[McLure 
\& Dunlop(2004)]{2004MNRAS.352.1390M} McLure, R.~J., \& Dunlop, J.~S.\ 2004, \mnras, 352, 1390

\bibitem[McLure 
\& Jarvis(2004)]{McLure2004b} McLure, R.~J., \& Jarvis, M.~J.\ 2004, \mnras, 353, L45

\bibitem[Moran et al.(1995)]{1995PNAS...9211427M} Moran, J., Greenhill, L., 
Herrnstein, J., Diamond, P., Miyoshi, M., Nakai, N., 
\& Inque, M.\ 1995, Proceedings of the National Academy of Science, 92, 11427

\bibitem[Narayan 
\& McClintock(2008)]{2008NewAR..51..733N} Narayan, R., \& McClintock, J.~E.\ 2008, \nar, 51, 733

\bibitem[Osterbrock 
\& Ferland(2006)]{Osterbrock2006agnbook} Osterbrock, D.~E., \& Ferland, G.~J.\ 2006, Astrophysics of gaseous nebulae and active galactic nuclei, 2nd.~ed.~ (Sausalito, CA: University Science Books) 

\bibitem[Peterson et al.(2004)]{Peterson2004} Peterson, B.~M., et 
al.\ 2004, \apj, 613, 682

\bibitem[Rafter(2010)]{Rafter2010} Rafter, S.~E. 2010, Ph.D. thesis, Georgia
State Univ.

\bibitem[Rafter, Crenshaw \& Wiita (2009)]{Rafter2009} Rafter, S.~E., Crenshaw, 
D.~M., \& Wiita, P.~J.\ 2009, \aj, 137, 42 (Paper I)

\bibitem[Saripalli 
\& Subrahmanyan(2009)]{2009ApJ...695..156S} Saripalli, L., \& Subrahmanyan, R.\ 2009, \apj, 695, 156

\bibitem[Schoenmakers et al.(2000)]{2000MNRAS.315..371S} Schoenmakers, 
A.~P., de Bruyn, A.~G., R{\"o}ttgering, H.~J.~A., van der Laan, H., 
\& Kaiser, C.~R.\ 2000, \mnras, 315, 371

\bibitem[Sikora et al.(2007)]{Sikora2007} Sikora, M., Stawarz, 
{\L}., \& Lasota, J.-P.\ 2007, \apj, 658, 815

\bibitem[Spergel et al.(2003)]{Spergel2003} Spergel, D.~N., et al.\ 
2003, \apjs, 148, 175

\bibitem[Urry 
\& Padovani(1995)]{1995PASP..107..803U} Urry, C.~M., \& Padovani, P.\ 1995, \pasp, 107, 803 

\bibitem[White et al.(2007)]{2007ApJ...654...99W} White, R.~L., Helfand, 
D.~J., Becker, R.~H., Glikman, E., \& de Vries, W.\ 2007, \apj, 654, 99

\end{thebibliography}
\end{document}